\documentclass[twocolumn,aps,prl,showpacs,superscriptaddress,floatfix]{revtex4}
\usepackage{graphicx}
\usepackage{times}
\usepackage{nicefrac}
\usepackage{amsmath}
\usepackage{amsfonts}
\usepackage{amssymb}
\usepackage{amsthm}
\usepackage{epsf}
\usepackage{bm}
\usepackage{bbm}
\usepackage{times}

\usepackage{dcolumn}
\newcolumntype{.}{D{x}{}{-1}}

\def\corresponds{{\lower.2ex\hbox{=}}{\rm\kern-.75em^\triangle}}
\def\succsim{\succ\kern-.9em_\sim\kern.3em}
\def\precsim{\prec\kern-1em_\sim\kern.3em}
\def\slantfrac#1#2{\kern1em^{#1}\kern-.3em/\kern-.1em_{#2}}
\def\lfrac#1#2{{}^{#1\!}\kern-.0em/_{#2}}

\def\buildrel#1\under#2{\mathrel{\mathop{\kern0pt #2}\limits_{#1}}}

\newcommand{\Za}{Z \alpha}

\begin{document}

\title{Nonrelativistic QED approach to the bound-electron $\bm{g}$ factor}

\author{Krzysztof Pachucki}
\affiliation{Institute of Theoretical Physics, Warsaw University,
ul.~Ho\.{z}a 69, 00--681 Warsaw, Poland}

\author{Ulrich D.~Jentschura}
\affiliation{Max--Planck--Institut f\"ur Kernphysik,
Saupfercheckweg 1, 69117 Heidelberg, Germany}

\author{Vladimir A. Yerokhin}
\affiliation{Physics Dept., St. Petersburg State University,
Oulianovskaya 1, Petrodvorets, St. Petersburg 198504, Russia}

\begin{abstract}

Within a systematic approach based on nonrelativistic quantum electrodynamics
(NRQED), we derive the one-loop self-energy correction of order $\alpha
(\Za)^4$ to the bound-electron $g$ factor. In combination with numerical
data, this analytic result improves theoretical predictions for the
self-energy correction for carbon and oxygen by an order of magnitude. Basing
on one-loop calculations, we obtain the logarithmic two-loop
contribution of order $\alpha^2 (\Za)^4 \ln[(Z\alpha)^{-2}]$ and the
dominant part of the corresponding constant term. The results obtained
improve the accuracy of the theoretical predictions for the $1S$
bound-electron $g$ factor and influence the value of the electron mass
determined from $g$ factor measurements.

\end{abstract}


\maketitle

There has been significant progress in experimental investigations of the
bound-electron $g$ factor during recent years~\cite{haeffner:00:prl,verdu:04}.
The $g$ factor value in these measurements is obtained
from the ratio of the electronic Larmor precession
frequency $\omega_{\rm L}$ and the cyclotron frequency of the ion in the trap
$\omega_{\rm c}$, according to
\begin{equation}
\frac{\omega_{\rm L}}{\omega_{\rm c}} =
\frac{g}{2}\, \frac{|e|}{q}\, \frac{m_{\rm ion}}{m}\,,
\label{1a}
\end{equation}
where  $e$ is the elementary charge, $q$ is the charge of the ion, $m_{\rm
ion}$ is the ion mass, and $m$ is the electron mass. The present accuracy
of the experimental results for carbon and oxygen is already below the 
level of 1 part per billion
and is likely to be improved in the near future. In order to match the
experimental precision achieved for H-like ions, accurate calculations of the
one-loop self-energy \cite{blundell:97:pra,persson:97:g,
yerokhin:02:prl,%
yerokhin:04:pra}, vacuum-polarization \cite{persson:97:g,karshenboim:02:plb},
and nuclear-recoil \cite{shabaev:01:rec,%
shabaev:02:prl} corrections have been performed during the last decade. An
important result of these studies is the possibility to extract the electron
mass from the experimental value for $\omega_L/\omega_c$ according to
Eq.~(\ref{1a}). Such a determination was presented in
Refs.~\cite{beier:02:prl,yerokhin:02:prl,verdu:04} based on the experimental result
for H-like carbon \cite{haeffner:00:prl} and oxygen. It provided an improvement of the
accuracy of the electron mass  by a factor of 4, as compared to the
previous value based on measurements involving protons and electrons in Penning traps
\cite{farnham:95:prl}. As a result, the presently recommended value for the electron
mass \cite{CODATA02} is derived mainly from the bound-electron $g$ factor.

The current uncertainty of the theoretical values for the $g$ factor in
H-like carbon and oxygen (as well as for other low- and medium-$Z$ ions)
originates predominately from the two-loop binding QED correction. This
correction is presently known \cite{grotch:70:prl} only to its leading order
in $\Za$, namely $\alpha^2 (\Za)^2$, the corresponding contribution being of pure
kinematical origin. The next correction enters in order $\alpha^2 (\Za)^4$
and results from non-trivial two-loop binding QED effects that have not been
addressed theoretically up to now.

The goal of the present investigation is to formulate an approach for the
systematic derivation of higher-order QED corrections to the $g$ factor of a
bound electron. Applicability of this approach is demonstrated for the
one-loop self-energy correction, for which direct numerical calculations to
all orders in $\Za$ are available. The analytical contribution derived here
is used in
order to reduce the uncertainty of the numerical results for carbon and
oxygen by an order of magnitude, this being the second largest error of the
corresponding theoretical values for the $g$ factor \cite{yerokhin:02:prl}.
The developed method is then applied to the evaluation of the two-loop
self-energy correction. We derive the complete result for the logarithmic
contribution of order $\alpha^2(\Za)^4 \ln[(Z\alpha)^{-2}]$ and a large part
of the corresponding constant term. The results obtained improve the accuracy of
the theoretical values for the $1S$ bound-electron $g$ factor in carbon and
oxygen and influence the electron-mass determination derived from the
corresponding experimental results.

The computational approach is based on the Dirac-Coulomb Hamiltonian
that is modified by the presence of the free-electron form
factors $F_1$ and $F_2$
\cite{itzykson:book},
\begin{eqnarray}
\label{HDm}
H &=& \bm{\alpha} \cdot
\left[\bm{p} - e \, F_1(\Delta) \, \bm{A}\right]
+ \beta\,m + e\,F_1(\Delta) \, A_0
\nonumber \\ && +
F_2(\Delta) \, \frac{e}{2\,m} \, \left({\mathrm i}\,
\bm{\gamma} \cdot
\bm{E} - \beta \, \bm{\Sigma} \cdot \bm{B} \right)\,,
\end{eqnarray}
where $\Delta$ denotes the Laplacian operator. This Hamiltonian accounts only partly for
the interaction of the electron with high-frequency photons.
The remaining high-energy contribution can be represented as a local type of
effective interaction. It is obtained by matching the low-energy limit of
scattering amplitudes derived from the Hamiltonian (\ref{HDm}) and from full
QED in a way that will be discussed below.

In the calculation of electron self-energy corrections, it is often
convenient to evaluate contributions due to the high- and the low-energy
virtual photons separately. The separation is achieved by introducing a
certain cut-off parameter, which in our calculation is chosen to be the photon
mass.
The low-energy part is then identified as the difference of the
contributions with the massless and the massive photons. In
order to calculate it, we perform the Foldy-Wouthuysen transformation of the
Hamiltonian $H$ followed by the Power-Zienau transformation, as described in
Ref.~\cite{pachucki:04:pra},
\begin{eqnarray}
\lefteqn{H_{PZ} = \frac{\bm{p}^2}{2\,m} -\frac{Z\,\alpha}{r} - e\,\bm{r}\cdot\bm{E}
- \frac{e}{2\,m}\,(1+\kappa)\,\bm{\sigma}\cdot\bm{B}}
\nonumber \\ &&
- \frac{e}{2\,m}\,\bm{L}\cdot\bm{B} +
\frac{Z\,\alpha}{4\,m^2}\,(1+2\,\kappa)\,\frac{\bm{\sigma}\cdot\bm{L}}{r^3}
\nonumber \\ &&
+\frac{p^2}{4\,m^3}\,e\,\bm{\sigma}\cdot\bm{B}
+\frac{e\,\kappa}{4\,m^3}\,(\bm{p}\cdot\bm{\sigma})\,(\bm{p}\cdot\bm{B})
\nonumber \\ &&
-\frac{e\,(1+\kappa)}{2\,m}\,\sigma^i\,r^j\,\partial_j B^i
-\frac{e\,(1+2\,\kappa)}{4\,m^2}\,\bm{\sigma}\cdot\bm{E}\times\bm{p}
\nonumber \\ &&
-\frac{e\,(1+2\,\kappa)}{8\,m^2}\,
(\bm{\sigma}\times\bm{E})\cdot(\bm{r}\times\bm{B})
\nonumber \\ &&
-\frac{e\,(1+2\,\kappa)}{8\,m^2}\,\frac{Z\,\alpha}{r^3}\,
(\bm{r}\times\bm{\sigma})\cdot(\bm{r}\times\bm{B})\,.
\label{01}
\end{eqnarray}
Here, $\bm{E} =\left. \bm{E}\right|_{\bm{r}=\bm{0}}$, $\bm{B}
=\left.\bm{B}\right|_{\bm{r}=\bm{0}}$, and $\kappa \equiv F_2(0)$.
In the Hamiltonian $H_{PZ}$ we neglected the spin independent terms
and the $\Delta$-dependence of the form factors.
Moreover, the terms with $\kappa$  will be needed only in
the two-loop calculation.

We consider now the one-loop self-energy contribution to the bound-electron
$g$ factor. It is represented by the sum of 3 parts, $ \delta g^{(1)} =
g_1^{(1)} + g_2^{(1)} + g_3^{(1)}\,$. The first part comes from the
free-electron form factors in the Hamiltonian $H$, the second part is due to an
additional term that matches scattering amplitudes, and the third part is a
low-energy-photon contribution that originates from the Hamiltonian $H_{PZ}$,
and is very similar to the Bethe logarithm in the hydrogen Lamb shift. All
these parts are calculated in the following.

{\em Form-factor contribution.}---We evaluate this part by separating the
Hamiltonian (\ref{HDm}) into the unperturbed (Dirac) Hamiltonian and the
interaction part and applying the standard Rayleigh-Schr\"odinger
perturbation theory. Taking into account that only the first 2 terms in the
$\Delta$ expansion of form factors contribute to the order of interest, we
write the interaction Hamiltonian as
\begin{eqnarray}
\delta H &=& \frac{e\,\kappa}{2\,m}\,\Bigl({\rm i}\,\bm{\gamma}\cdot\bm{E} -
\beta \,\bm{\Sigma}\cdot\bm{B}\Bigr)
 \nonumber \\ &&
  + e\,F'_1(0)\,\Delta A_0
+\frac{{\rm i}\,e}{2\,m}\, F'_2(0)\Delta(\bm{\gamma}\cdot\bm{ E}) \,,
\label{03}
\end{eqnarray}
where the slope of the form factors are known to be $F'_1(0) =
\frac{\alpha}{2\,\pi}\, (-\nicefrac{1}{4}- \nicefrac{2}{3}\,\ln \mu)$ and
$F'_2(0) = \alpha/(12\,\pi)$, where $\mu$ is the ratio of the photon mass to the
electron mass. Applying perturbation theory in first
and second orders and separating contributions linear in the
magnetic field, we obtain the correction to the $g$ factor of an $nS$ state,
\begin{equation}
g_1^{(1)} = \frac{\alpha}{\pi}\biggl[ 1+\frac{(Z\,\alpha)^2}{6\,n^2} -
\frac{(Z\,\alpha)^4}{n^3}\,\left(\frac76
  +\frac5{24n} +\frac{16}{3}\, \ln \mu \right) \biggr] \,.
\label{11}
\end{equation}

{\em Spin-dependent scattering amplitude.}---It
represents a high-energy contribution which goes beyond the on-shell
form-factor treatment. Here we only sketch the idea of the derivation; the
details will be presented elsewhere. We first introduce the skeleton
amplitude of the free electron scattering off both the Coulomb and the
magnetic external field. Then we add an electron self-energy loop inserted
into the skeleton diagram in all possible ways. Infrared divergences present
in loop-momentum integration are regularized by employing the photon mass.
Next, we subtract the skeleton amplitude with vertices modified by the
electron form factors $F_1$ and $F_2$, expand in all external momenta, and
keep terms of the third power in external momenta only. The resulting
amplitude can be represented by the following effective-interaction
Hamiltonian ($\bm{B}=\, $const.)
\begin{equation}
\delta H_2 = \frac{\alpha}{\pi}\,\biggl[ \frac{e^2}{4}\,\sigma_i B_i\,\partial_j
E_j + \frac{e^2}{3}\,\ln\mu\;\sigma_i\,B_j\,\partial_i\,E_j \biggr]\,. \label{12}
\end{equation}
The corresponding correction to the
energy reads
\begin{equation}
\delta E_2^{(1)} =
\langle\phi|\left(-Z\,\alpha^2\right)\,\delta^3(r)\,e\,\bm{\sigma}\cdot\bm{B}\,
\Bigl(1+\frac{4}{9}\,\ln\mu\Bigr)|\phi\rangle\,, \label{13}
\end{equation}
which induces the following contribution to the $g$ factor
\begin{equation}
g_2^{(1)} = \frac{\alpha}{\pi}\,\frac{(Z\,\alpha)^4}{n^3}\,
\Bigl(4+\frac{16}{9}\,\ln\mu\Bigr)\,. \label{14}
\end{equation}

{\em Low-energy part.}---This contribution is induced by the virtual photon of
low energy. We first recall the expression for the low-energy part of the
Lamb shift,
\begin{eqnarray}
\delta E_L &=& \frac{2\,\alpha}{3\,\pi}\,\int {\rm d}k \,k^3\,
\langle\phi|\bm{r}\frac{1}{E-H_S-k}\,\bm{r}|\phi\rangle \nonumber  \\ &=&
\frac{\alpha}{\pi}\,\frac{(Z\,\alpha)^4}{n^3}\,\frac{4}{3}\,\biggl[
  \ln \frac{\mu}{(Z\,\alpha)^2}+\frac56-\ln k_0\biggr]\,, \label{16}
\end{eqnarray}
where
$H_S$ is the Schr\"odinger Hamiltonian. In the above, we assume the implicit
difference between massless and massive photons and keep only the terms that
do not vanish when $\mu \to 0$. In practice, one calculates this with a cut-off
$k<\epsilon$ and later performs the replacement $\ln 2\epsilon \to \ln
\mu+\nicefrac56$ \cite{itzykson:book}. We mention that this replacement is not
unique and its specific form depends on the actual integrand.
Eq.~(\ref{16}) coincides with the standard
definition of the Bethe logarithm $\ln k_0$, which has the explicit values $\ln
k_0(1S) = 2.984~128~555$ and $\ln k_0(2S) = 2.811~769~893$.

Here, we are interested in all possible relativistic corrections to
Eq.~(\ref{16}) induced by the Hamiltonian $H_{PZ}$ which are linear in the
$\bm{B}$ field. There are 6 such corrections presented in Table I.  The terms
involving $\bm{E}$ and $\partial_j B^i$ represent corrections to the vertex
$\left(-e\,\bm{r}\cdot\bm{E}\right)$, and the others yield
corrections to $H$, $E$, and $\phi$. The results listed in the third column
of Table I involve the standard Bethe logarithm $\ln k_0$ and its
modification $\ln k_3$ defined as
\begin{eqnarray}
\int {\rm d} k\, k^2\,
\langle\phi|\bm{r}\,\frac{1}{E-H_S-k}\,\frac{1}{r^3}\,\frac{1}{E-H_S-k}\,\bm{r}\,|\phi\rangle
  \nonumber \\
= -4\,\frac{(\Za)^3}{n^3} \left[\ln \frac{\mu}{(Z\,\alpha)^2}+\frac56-\ln
  k_3\right].\ \  \label{17}
\end{eqnarray}
We calculate $\ln k_3$ numerically,  using a finite-difference representation
of the Schr\"odinger Hamiltonian $H_S$, and obtain the following results for
the $1S$ and $2S$ states: $\ln k_3(1S) = 3.272~806~545$ and $\ln k_3(2S) =
3.546~018~666$. Finally, we present the sum of all 6 contributions from Table
I as
\begin{equation}
g_3^{(1)} = \frac{\alpha}{\pi}\,\frac{(Z\,\alpha)^4}{n^3}\,\frac{32}{9}\,
\biggl[\ln\frac{\mu}{(Z\,\alpha)^2} -\frac{5}{12} -\frac{\ln k_0}{4}
-\frac{3}{4}\,\ln k_3 \biggr].\label{26}
\end{equation}

\begin{table*}
\begin{center}
\begin{minipage}{14.3cm}
\begin{center}
\caption{\label{tab:tab1} Breakdown of the low-energy contribution to the
bound-electron $g$ factor.}
\begin{ruledtabular}
\begin{tabular}{rrrc}
   \multicolumn{1}{c}{\#}
& \multicolumn{1}{c}{$\delta H$} & \multicolumn{1}{c}{$\delta g$} &
\multicolumn{1}{c}{two-loop prefactor} \\ \colrule 1   &
$\frac{p^2}{2m}\,\frac{e\,\bm{\sigma}\cdot\bm{B}}{2\,m^2}$ &
$\frac{8}{3}\,\Bigl[-\frac{1}{6}-\ln k_0 - \ln(Z\,\alpha)^2+\ln\mu\Bigr]$ &
${\kappa}/{3}$ \\ 2   &
$-\frac{e}{8\,m^2}\,(\bm{r}\times\bm{\sigma})\cdot(\bm{r}\times\bm{B})$ &
$-\frac{16}{9}\,\Bigl[ \frac{1}{3}-\ln k_0 - \ln(Z\,\alpha)^2+\ln\mu \Bigr]$
& $2\,\kappa$ \\ 3   &
$-\frac{e}{8\,m^2}\,(\bm{\sigma}\times\bm{E})\cdot(\bm{r}\times\bm{B})$ &
$\frac{8}{9}\,\Bigl[ \frac{5}{6}-\ln k_0 - \ln(Z\,\alpha)^2+\ln\mu \Bigr]$ &
$2\,\kappa$ \\ 4   & $-\frac{e}{4\,m^2}\,\bm{\sigma}\cdot\bm{E}\times\bm{p}-
\frac{e}{2\,m}\,\bm{L}\cdot\bm{B}$ & $\frac{8}{3}\,\Bigl[ \frac{1}{2}-\ln k_0
- \ln(Z\,\alpha)^2+\ln\mu \Bigr]$ & $2\,\kappa$ \\ 5   &
$-\frac{e}{2\,m}\,\sigma^i\,r^j\,\partial_j B^i- \frac{e}{2\,m}\,\bm{L}\cdot\bm{B}$
& $-\frac{32}{9}\,\Bigl[ \frac{13}{12}-\ln k_0 - \ln(Z\,\alpha)^2+\ln\mu
\Bigr]$ & $\kappa$ \\ 6   &
$\frac{Z\,\alpha}{4\,m^2}\,\frac{\bm{\sigma}\cdot\bm{L}}{r^3} -
\frac{e}{2\,m}\,\bm{L}\cdot\bm{B}$ & $\frac{8}{3}\,\Bigl[ \frac{1}{2}-\ln k_3
- \ln(Z\,\alpha)^2+\ln\mu \Bigr]$ & $2\,\kappa$ \\
\end{tabular}
\end{ruledtabular}
\end{center}
\end{minipage}
\end{center}
\end{table*}

{\em Two-loop contribution.}---The complete two-loop
self-energy of order $(\Za)^4$ consists of contributions related to the
electron form factors, to the two-photon scattering amplitude, and to the
low-energy part. In the present investigation we derive an expression for
the low-energy part only. This gives the complete result for the
logarithmic contribution of relative order $(\Za)^4$. Moreover, we
observe that in the one-loop case, the low-energy part yields the dominating
contribution of about 75\% of the constant term. Arguably, this is a general
feature of all radiative corrections, another example being the hydrogen Lamb
shift. We thus assume that also in the two-loop case, the low-energy part
provides the dominant contribution to the constant term. A derivation of the
remaining two-loop contributions can in principle be carried out along the lines presented
above.

Let us now identify the two-loop low-energy correction. When {\em both}
photons are of low energy, the $\bm{B}$-dependent part is of a higher order
and thus negligible. The contribution of interest comes when only {\em one}
photon is of low energy. The second photon effectively modifies the vertex,
and only the part with the anomalous magnetic moment is relevant, as the
slope of the form factors contributes to higher orders.
There are two equivalent contributions obtained by interchanging the
photons, which results in an additional factor of 2.
The effective Hamiltonian that accounts for the anomalous magnetic
moment is given by Eq.~(\ref{01}). The calculation of the Bethe-logarithmic
contributions is the same as for the one-loop case, but involves different
overall factors for each term, which are listed in the fourth column of Table
I. Using the one-loop results, we obtain for the sum of all low-energy
contributions,
\begin{eqnarray}
g_3^{(2)} &=& \left(\frac{\alpha}{\pi}\right)^2\,\frac{(Z\,\alpha)^4}{n^3}\,
   \left[ \frac{56}{9}\, \ln \frac{\mu}{(Z\alpha)^{2}}
\right.  \nonumber \\ && \left.
+\frac{44}{27} -\frac89\,\ln k_0 -\frac{16}{3}\,\ln k_3
              \right] \,,
\label{27}
\end{eqnarray}
where the numerical value of the constant term is $a_{40}^{(2)}(1S) =
-18.477\,948\,664\,(1)$ and $a_{40}^{(2)}(2S) = -19.781\,820\,939\,(1)$. The
term with $\ln\mu$ is canceled by corresponding contributions coming from
the slope of the form factors and the two-loop scattering amplitude.

\begin{table*}
\begin{center}
\begin{minipage}{16.0cm}
\caption{Individual contributions to the $1s$ bound-electron
$g$ factor, $1/\alpha$ from \cite{CODATA02} is $137.035\,999\,11(46)$.
\label{tablegfact} }
\begin{ruledtabular}
\begin{tabular}{l..}
& \multicolumn{1}{c}{$^{12}{\rm C}^{5+}$ }
& \multicolumn{1}{c}{$^{16}{\rm O}^{7+}$}\\
\hline
Dirac value (point)             &  1x.998$\,$ 721$\,$ 354$\,$ 39$\,$(1)  & 1x.997$\,$ 726$\,$ 003$\,$ 06$\,$(2) \\
Finite nuclear size             &  0x.000$\,$ 000$\,$ 000$\,$ 41         & 0x.000$\,$ 000$\,$ 001$\,$ 55 \\
Free QED,  $\sim (\alpha/\pi)$  &  0x.002$\,$ 322$\,$ 819$\,$ 47$\,$(1)  & 0x.002$\,$ 322$\,$ 819$\,$ 47$\,$(1) \\
Binding SE, $\sim (\alpha/\pi)$ &  0x.000$\,$ 000$\,$ 852$\,$ 97         & 0x.000$\,$ 001$\,$ 622$\,$ 67$\,$(1) \\
Binding VP, $\sim (\alpha/\pi)$ & -0x.000$\,$ 000$\,$ 008$\,$ 51         &-0x.000$\,$ 000$\,$ 026$\,$ 37$\,$(1) \\
Free QED, $\sim (\alpha/\pi)^2 \cdots (\alpha/\pi)^4$
                                & -0x.000$\,$ 003$\,$ 515$\,$ 10         &-0x.000$\,$ 003$\,$ 515$\,$ 10        \\
Binding QED, $\sim (\alpha/\pi)^2(Z \alpha)^2$
                                & -0x.000$\,$ 000$\,$ 001$\,$ 13         &-0x.000$\,$ 000$\,$ 002$\,$ 01        \\
Binding QED, $\sim (\alpha/\pi)^2(Z \alpha)^4$
                                &  0x.000$\,$ 000$\,$ 000$\,$ 41$\,$(11) & 0x.000$\,$ 000$\,$ 001$\,$ 06$\,$(35) \\
Recoil                          &  0x.000$\,$ 000$\,$ 087$\,$ 63         & 0x.000$\,$ 000$\,$ 116$\,$ 97        \\
Total                           &  2x.001$\,$ 041$\,$ 590$\,$ 52$\,$(11) & 2x.000$\,$ 047$\,$ 021$\,$ 28$\,$(35)   \\
\end{tabular}
\end{ruledtabular}
\end{minipage}
\end{center}
\end{table*}

{\em Results and discussion.}---We first summarize our calculation for the
{\it one-loop} self-energy correction. The total analytic result
for an $nS$ state is 
\begin{eqnarray}
&& \delta g^{(1)} = \frac{\alpha}{\pi} \left\{1 + \frac{(Z\alpha)^2 }{6 n^2}
+ \frac{(Z\alpha)^4}{n^3} \left[\frac{32}{9} \ln[(Z\alpha)^{-2}] +
\frac{73}{54} \right. \right. \nonumber \\ && \left. \left. - \frac{5}{24 n}
- \frac{8}{9} \ln k_0 - \frac{8}{3} \ln k_3 \right] + (Z \alpha)^5
G_{n}(Z)\right\},\label{28}
\end{eqnarray}
where the remainder function $G_{n}(Z)$ incorporates all contributions of
higher orders in $\Za$, and the numerical value of  the constant term in order
$(\Za)^4$ is $a_{40}^{(1)} = -10.236~524~318(1)$ for the $1S$ state and
$a_{40}^{(1)} = -10.707~715~607(1)$ for the $2S$ state. The first two terms in
Eq.~(\ref{28}) are well known; the first one is the famous Schwinger
correction and the second was derived previously for the $1S$ state in
Ref.~\cite{grotch:70:prl}.

By subtracting all known terms of the $\Za$ expansion in Eq.~(\ref{28}) from
numerical data \cite{yerokhin:02:prl,yerokhin:04:pra}, one can isolate the
one-loop self-energy remainder $G_{n}(Z)$ and improve its numerical accuracy
for carbon and oxygen by extrapolating results for higher values of $Z$. The
higher-order contribution extracted directly from numerical results of
Ref.~\cite{yerokhin:02:prl} reads $G_1(6) = 22.19(24)$ and $G_1(8) =
21.86(6)$. An extrapolation of numerical data \cite{yerokhin:02:prl} for
$Z>8$ yields the results for the self-energy remainder  
$G_1(6) = 22.160(10)$ and $G_1(8) = 21.859(4)$, which are
significantly more accurate. In addition, we obtain the following result for
the total contribution of order $(\Za)^5$: $G_1(0) = 23.0$.

The result for the {\it two-loop} self-energy contribution is given by
Eq.~(\ref{27}). We estimate the uncertainty due to
uncalculated parts $g_1^{(2)}$ and $g_2^{(2)}$ as 30\% of the constant term.  
Explicitly, the two-loop self-energy correction for 
the $1S$ state is
\begin{equation}
\delta g^{(2)} = \Bigl(\frac{\alpha}{\pi}\Bigr)^2 \, (Z\alpha)^4 \,
\Bigl\{\frac{56}{9}\,\ln[(Z\alpha)^{-2}] - 18.5(5.5)\Bigr\}\,.
        \label{28a}
\end{equation}

We now turn to the experimental consequences of our calculation. A previous
compilation of theoretical contributions to the $1S$ bound-electron $g$ factor
was given in Ref.~\cite{yerokhin:02:prl}. In the present work we modify it in
several ways, with the corresponding contributions listed in Table
\ref{tablegfact}: {\em (i)} we employ the new, more accurate results for the
one-loop self-energy remainder; {\em (ii)} we use the analytic result of
Ref.~\cite{karshenboim:02:plb} for the leading term of the $\Za$ expansion of
the first-order magnetic-loop vacuum-polarization correction; {\em (iii)} we
include the leading part of the two-loop self-energy correction of order
$\alpha^2 (\Za)^4$ obtained in this work [Eq. (\ref{28a})]. We assume that
the uncertainty due to other uncalculated two-loop corrections 
is absorbed into the error bars of the constant term in
Eq.~(\ref{28a}).

As compared to the previous compilation
\cite{yerokhin:02:prl}, the accuracy of the theoretical value for carbon is
improved by a factor of 3. In case of oxygen, only a small improvement of
accuracy is achieved, but the theoretical value is shifted slightly outside
of the error bars given in Ref.~\cite{yerokhin:02:prl}.
The described modification of the theoretical predictions for the
bound-electron $g$ factor influences the electron-mass values derived from
the experiments on carbon \cite{haeffner:00:prl} and oxygen \cite{verdu:04}.
Following Refs.~\cite{beier:02:prl,verdu:04} and using the $g$ factor values
from Table~\ref{tablegfact}, we obtain the following results for the electron
mass,
\begin{eqnarray}
m(^{12}{\rm C}^{5+}) &=& 0.000\,548\,579\,909\,41\, (29)(3)\ {\rm u} \,, \\
m(^{16}{\rm O}^{7+}) &=& 0.000\,548\,579\,909\,87\, (41)(10)\ {\rm u} \,,
\end{eqnarray}
where the first uncertainty originates from the experimental
value for the ratio  $\omega_L/\omega_c$, and the second error comes from the
theoretical result for the bound-electron $g$ factor.

In summary, we have presented an approach for a systematic derivation of
higher-order QED corrections to the $g$ factor of a bound electron. We
obtained the complete result for the one-loop self-energy
correction of order $\alpha (\Za)^4$. The derived contribution is
in excellent agreement with the previous numerical calculation.
The developed approach was then applied to the most problematic two-loop
self-energy correction. We obtained the logarithmic contribution to order $\alpha^2
(\Za)^4\ln (\Za)^{-2}$ and the dominant part of the corresponding
constant term. As a result, we improved the accuracy of the theoretical
predictions for the $1S$ bound-electron $g$ factor for carbon and oxygen and
presented new values for the electron mass derived from the corresponding
measurements.

We wish to thank P. J. Mohr, W. Quint, T. Beier and V. Shabaev for helpful
conversations. This work was supported by EU grant No. HPRI-CT-2001-50034,
by RFBR grant No. 04-02-17574, and by a grant of the foundation "Dynasty".

\end{document}